\documentclass[a4paper,english,aps,prl,twocolumn,floatfix,showpacs,amsfonts,amssymb,superscriptaddress]{revtex4} 
\usepackage[T1]{fontenc}
\usepackage[latin1]{inputenc}
\usepackage{graphics}
\usepackage{amssymb}
\usepackage{amsmath}
\usepackage{overpic}

\newcommand{\be}{\begin{equation}}
\newcommand{\ee}{\end{equation}}
\newcommand{\bea}{\begin{eqnarray}}
\newcommand{\eea}{\end{eqnarray}}

\def\d{\delta}

\def\pd{\partial}
\def\nb{\nabla}

\def\br{{\bf r}}

\def\bA{{\bf A}}

\def\lb{\label}
\def\pref#1{(\ref{#1})}

\newcount\bozza \bozza=0
\ifnum\bozza=1
\newdimen\shift \shift=-2truecm
\def\lb#1{%
{\label{#1}\rlap{\kern\shift{$\scriptstyle#1$}}}}
\else\def\lb#1{\label{#1}} \fi

\begin{document}
\title{Superfluid density  and phase relaxation in superconductors
  with strong disorder}

\author{G.~Seibold} 
\affiliation{Institut F\"ur Physik, BTU Cottbus, PBox 101344, 03013 Cottbus,
Germany}
\author{L.~Benfatto} 
\affiliation{ISC-CNR and Department of Physics, University of Rome ``La
  Sapienza'',\\ Piazzale Aldo Moro 5, 00185, Rome, Italy}
\author{C.~Castellani}
\affiliation{ISC-CNR and Department of Physics, University of Rome ``La
  Sapienza'',\\ Piazzale Aldo Moro 5, 00185, Rome, Italy}
\author{J.~Lorenzana}
\affiliation{ISC-CNR and Department of Physics, University of Rome ``La
  Sapienza'',\\ Piazzale Aldo Moro 5, 00185, Rome, Italy}

\date{\today}

\begin{abstract}
As a prototype of a disordered superconductor we consider the attractive 
Hubbard model with on-site disorder. We solve the Bogoljubov-de-Gennes 
equations on two-dimensional finite clusters at zero temperature and 
evaluate the electromagnetic response to a vector potential. We find that 
the standard decoupling  between transverse and longitudinal response 
does not apply in the presence of disorder. Moreover the superfluid density 
is strongly reduced by the relaxation of the phase of the order parameter 
already at mean-field level when disorder is large. We also find that the 
anharmonicity  of the phase fluctuations is strongly enhanced 
by disorder. 
Beyond mean-field, this provides an enhancement of quantum fluctuations 
inducing a zero-temperature transition to a non-superconducting 
phase of disordered preformed pairs. 
Finally, the connection of our findings with the glassy physics for extreme
dirty superconductors is discussed.
\end{abstract}


\maketitle

In the last few years renewed interest emerged in the behavior of
disordered superconductors at the verge of the metal-insulator
transition. 
On the experimental side recent studies of
disordered superconducting films at low-temperatures
\cite{sacepe09,mondal10} have revealed the
existence of a striking ``pseudogap'' behavior at large disorder, 
with the tunneling conductance at zero bias starting to develop a
suppression at temperatures much larger than $T_c$. These
findings suggest quite generically a separation between the
energy scales associated to local pairing (gap and pseudogap) 
and superconducting phase
coherence (superfluid density) with different dependencies on disorder. 
In particular, it is this second energy scale (to be more specific, 
the superfluid stiffness) which is expected to control the 
stability of the superconducting phase in 
the proximity to the superconductor-insulator (SC-I) transition.

From the theoretical point of view the study of disordered
superconductors near the SC-I transition\cite{feigelman10} has been 
based either on a bosonic approach\cite{fisher90}, where the role of
phase fluctuations emerges naturally, or on a more microscopic 
fermionic approach\cite{randeria01,feigelman07,dubi07,randeria10}, 
which has put the focus on the emergence of a short-scale
inhomogeneity induced by the strong disorder. In particular, in the
latter case the appearance of two characteristics
features of the disordered SC state has been demonstrated: 
the spontaneous emergence of spatial structures in the local pairing gap  
and a general suppression of the phase
coherence, which reflects in a decrease of the global superfluid
stiffness\cite{randeria01,dubi07,feigelman07,randeria10}.  
These results appear already at mean-field level, as it has been shown by the
Bogoljubov-de-Gennes (BdG) solution for a 2D SC system
in the presence of on-site disorder\cite{randeria01}, which serves
also as comparison for more refined Monte Carlo results including
thermal\cite{dubi07} or quantum\cite{randeria10} phase fluctuations.
Recently it has been argued that eventually a non self-averaging 
character typical of glassy physics will appear at very strong
disorder\cite{IMFIM}.

\begin{figure}[t]
\includegraphics[width=8cm,clip=true]{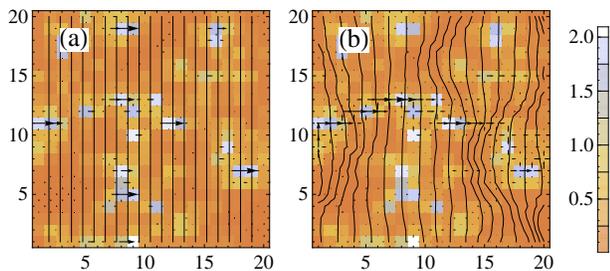}
\caption{(Color online) Distribution of the local current (arrows) and lines 
  of constant $\tilde \theta$ superimposed to the
  map of the order parameter, neglecting phase relaxation (a) and
  allowing for it (b). We applied a
  transverse field $A=0.003$, for a disorder amplitude $V_0/t$=2.
  Notice that in (b), while the percolative pattern of the current
  goes along the pattern of largest $\Delta_i$ values, the
  gauge-invariant phase gradient is larger along the
  locations of smaller $\Delta_i$ values.}
\label{fig-curr}
\end{figure}

In the presence of disorder a non-trivial problem is posed by the
correct computation of the superfluid stiffness $D_s$  and by the 
understanding of the processes leading to the supression of $D_s$. 
In the clean translational invariant case the
superfluid behavior of the system reflects the ``rigidity'' of
the phase against an applied transverse vector potential. This leads
to a purely diamagnetic response of the current, which is the hallmark
of the Meissner effect and the superfluid behavior of clean superconductors. At
small disorder a paramagnetic response appears and the superfluid
stiffness decreases with respect to the pure diamagnetic case. 
This paramagnetic current is the response of BdG quasiparticles 
to the external vector potential, while keeping   
the phase of the order parameter unchanged. 
Within the BdG approach, one can then account for this effect by
evaluating $D_s$ as a  
disorder average over the BCS response function (the BCS current-current 
correlation function with no vertex corrections), computed with the
BdG solution for the disordered system (cf. e.g. Ref. \cite{randeria01}). 
This procedure, which for small disorder is equivalent to the dirty-BCS limit, 
relies on the decoupling between longitudinal and transverse electromagnetic 
response: the phase of the order parameter does
(does not) react to a  longitudinal (transverse) field. However, this decoupling holds exactly only in clean systems.  Our main
results in this work are: 
{\em i)} For strong disorder we show that such a decoupling is strongly 
violated, leading to a dramatic 
decrease of the phase stiffness with respect to the dirty BCS case,
due to the additional paramagnetic suppression
coming from the phase relaxation to
the applied transverse vector potential. 
 {\em ii)} We discuss the behavior of this paramagnetic phase response in
connection to the formation of 
self-organized structures of the current  in real space 
at scales much larger than the lattice spacing 
and the SC coherence length
[cf. Fig. \ref{fig-curr}(b)], which share some analogies with the
glassy features discussed in Refs. \cite{IMFIM}. 
{\em iii)} We show that 
the anharmonic phase fluctuations become strongly enhanced already at
intermediate disorder with respect to the clean case, leading to a
sizeable increase of quantum corrections to the superfluid stiffness due to phase fluctuations, with respect to a clean system.
These results
can account for the recently observed
deviation of the superfluid stiffness from the dirty-BCS limit in NbN
films\cite{mondal10}, and offer new insight for the understanding 
of the SC-I transition.

Our starting hamiltonian is the attractive Hubbard model with
local disorder:
\begin{equation}
\lb{defh}
H=-t \sum_{\langle ij \rangle  \sigma}c^\dagger_{i\sigma}c_{j\sigma}+h.c. -
|U|\sum_{i}n_{i\uparrow}n_{i\downarrow} +\sum_{i\sigma}V_i
n_{i\sigma}, 
\end{equation}
which we solve using the BdG equations \cite{degennes} by allowing for
a site dependent SC order parameter
$\Delta_i=|U|\langle c_{i\downarrow}c_{i\uparrow}\rangle$. 
The first sum is over nearest-neighbors pairs
and we work on a $L\times L$ system  
using units such that lattice spacing and $\hbar=c=e=1$. The local potential $V_i$ is randomly 
distributed between $-V_0 \le V_i \le V_0$. 
We present results for density  $n=0.875$ and 
 at large SC coupling
  ($U/t=-5$), where the phase-relaxation effects and the connection to the
  glassy physics become more evident. Similar results are found in a
  wide parameter range.

The superfluid stiffness $D_s$ in  the $a=x,y$ direction 
(corresponding in the continuum limit to $D_s=n_s/m$ with
$n_s$ being the superfluid density) is defined as the static limit of the transverse part, $K_T$, of the
electromagnetic kernel, $K_{ab}$, which describes the current 
response, $j_a = K_{ab} A_b$, to an applied electromagnetic 
field $A_b$ (summation over repeated indices is
implicit)\cite{schrieffer,scalapino93}.

In principle, the correct evaluation of the superfluid stiffness requires 
the knowledge of the electromagnetic kernel $K_{ab}$ within a so-called
conserving approximation, i.e. an approximation that respects the
gauge invariance of the theory. As it is well know, the BCS
approximation for $K^{BCS}_{ab}$ is not conserving since, by neglecting 
the vertex corrections, it does not include 
the contribution of the phase relaxation to the current-current
correlation function 
\cite{schrieffer,scalapino93}. This is not a problem in the clean case
as  far as the transverse response is concerned, since phase fluctuations
contribute only to the longitudinal part of $K_{ab}$\cite{schrieffer}. 
It is easy to check that this no longer 
holds in the presence of disorder. Indeed, upon considering also the non-local 
character of the BCS kernel for a disordered system,
the effective action for phase fluctuations\cite{benfatto01} can in general be written at
Gaussian level as:
\be
\lb{sgdis}
S_g=\frac{1}{8}\int d^2\br d^2\br' (\nb \theta_\br-2\bA_\br)_a
K^{BCS}_{ab}(\br,\br')  (\nb \theta_{\br'}-2\bA_{\br'} )_b .
\ee
%
In the absence of disorder the above
expression simplifies considerably, since the BCS kernel 
$K^{BCS}_{ab}$ depends only on the difference $\br-\br'$, so that by making
the Fourier transform and approximating the BCS kernel with its (constant) 
long-wavelenght limit  $K^{BCS}_0\equiv K^{BCS}_{aa}(0)$, the action reduces to:
\be
\lb{sgclean}
S_g=\frac{1}{8}\int d^2 \br  K^{BCS}_0 (\nb \theta-2\bA)^2 .
\ee
Integrating by parts the term $\nb \theta\cdot \bA$, 
one finds that $\theta$ couples only with the {\em longitudinal} part
of the electromagnetic field. Integrating out the phase one finds
$D_s=K_T=K^{BCS}_0$ {\it i.e.} the stiffness is given by the BCS
kernel. Modeling disorder by a position dependent kernel in 
Eq.~(\ref{sgclean}) one sees  that the
integration by parts leads to a coupling of the phase with both the
longitudinal and the transverse part of the gauge field.  
The same is true if  Eq.~(\ref{sgdis}) is used which now depends
separately on $\br$ and $\br'$. For strong disorder one is not allowed to
substitute the BCS kernel in Eq.~\pref{sgdis} with its
space-disordered average which will restore translational invariance
and erroneously lead to the simplified Eq.~\pref{sgclean}.
The consequences of this transverse-longitudinal decoupling breakdown
are:  
(i) a change in the SC order-parameter phase even
in the presence of a transverse field, contrary to the clean case; (ii)
a failure of the BCS response function to compute the superfluid stiffness.

In order to put these arguments on a quantitative basis 
we first show explicitly the behavior of the current in the presence of a constant vector potential
$A$ in the $x$ direction. Using the Peierls substitution, this corresponds to a
change in the hopping term of Eq.\ \pref{defh}  along $x$:
\begin{equation}
\lb{txa}
T_x(A)=-t \sum_{i\sigma}\left[ e^{-iA} c_{i+x,\sigma}^\dagger c_{i\sigma}+h.c.\right].
\end{equation}
In a torus geometry (periodic boundary conditions) a constant $A$ cannot be gauged away, and it corresponds to a flux
$\Phi=AL$ through the torus.  For a given disorder
configuration the microscopic current along $x$ at finite $A$  
is then the sum of the paramagnetic and diamagnetic contribution
%
$j_x  =-\frac{\pd  T_x(A)}{\pd (A)}=
\langle j_x^P \rangle +\langle j_x^D \rangle$.
%
In the clean case $j_x$ can be directly derived
from Eq. \pref{sgclean}, and is proportional to
the gauge-invariant phase gradient $\nb \tilde \theta\equiv \nb
\theta -2\bA$. Here $\tilde \theta$ is the phase of the order parameter
if we would eliminate ${\bf A}$ by a gauge transformation. 
Since for an applied transverse field $\nb \theta=0$,
then $j_x$ reduces to the constant value $-AD_s$.

For the disordered system the result is radically different.  If one
  neglects phase relaxation (BCS), each piece of the system responds
  according to its local stiffness, which is essentially determined by
  the local order parameter [Fig.~\ref{fig-curr}(a)]. Because this
  solution is not a saddle point, the current  violates
  charge conservation and is clearly unphysical. 
Allowing for the phase to relax  [Fig.~\ref{fig-curr}(b)]
both the current and the gauge invariant phase gradient [proportional
to the line density in Fig.~\ref{fig-curr}(a)] are strongly space
dependent. This dependence is still correlated to the order parameter
but now it is also conditioned by the existence of a percolative path.
This means that the current response at one point can be strongly
influenced by the local stiffness at very far points. Notice that isolated
regions with a robust order parameter have practically zero current in
panel (b).  

The gauge-invariant phase gradient $\nb\tilde\theta$ is 
large predominantly in regions with minimum values 
of the order parameter $\Delta_i$. This can be understood by
mapping the problem to a random resistor network\cite{par98,kir73} where the bad
SC regions map into poor conducting regions and $\tilde
\theta$ maps into the electrostatic potential,
 so large ``potential'' drops concentrate in the ``poor'' conducting
 regions.  
Moreover, in contrast to the clean isotropic
case, where $j_x$ has the same direction than the gauge-invariant
$\nb \tilde \theta$ (see Eq.\ \pref{sgclean}), here one expects 
a tensorial relation between the two, as given in Eq.\ \pref{sgdis}
 and confirmed by the results of Fig.\ \ref{fig-curr}. It is worth
 stressing that while the 
local SC order parameter can be quite small, the local density of states
always shows a sizable gap, as found in various previous analyses\cite{randeria01,randeria10},
supporting the view that  coherent superconductivity takes place in
Fig.~\ref{fig-curr} on a system of localized preformed pairs. 

\begin{figure}[thb]
\includegraphics[scale=0.3,clip=true]{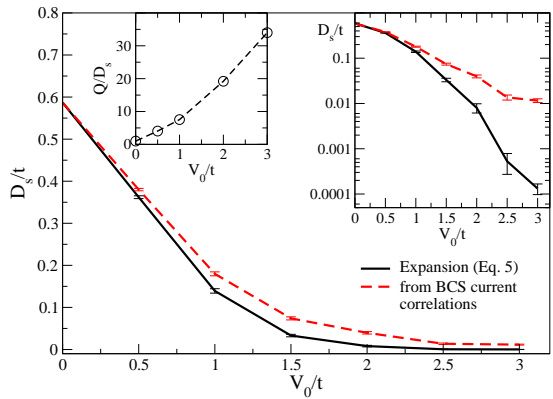}
\caption{(Color online) Main panel: Comparison between the superfluid stiffness 
  computed using Eq.\
  \pref{defda} and its BCS counterpart
  as a function of disorder.
  Upper right inset: Same data plotted on a log-log scale to 
  demonstrate increasing deviations for large $V_0/t$.
  Upper left inset: Ratio between anharmonic ($Q$) and quadratic coefficient
  ($D_s$) of the expansion Eq. \pref{enexp} (with $Q/D_s=1.01$ at
  $V=0$). Note that for the given parameter set $\xi_0 \approx 1a$ so
  that $J'/J \approx Q/D_s$. } 
\label{fig-ds}
\end{figure}

A direct procedure to compute the superfluid stiffness 
beyond BCS that accounts for the SC-phase relaxation, like those shown in Fig.~\ref{fig-curr}, 
is given by the change in the ground-state energy $E$ in
the presence of the constant vector potential of Eq.\ \pref{txa},\cite{khon,scalapino93} 
%
\be
\lb{defda}
{D_s}=\frac{1}{L^2}\frac{\pd ^2 E(A)}{\pd A^2}.
\ee  
%
The resulting values of $D_s$ are shown in Fig.\ \ref{fig-ds},
along with their BCS counterparts.
While the effects seem not dramatic at the scale of the bare stiffness they are gigantic
when the stiffness is small due to strong disorder, as can be seen in
the right inset.  Such a dramatic difference is clearly due to the 
global reorganization of the current, i.e. only good SC regions that are along the percolative
path contributes to the stiffness. In the BCS case instead the global
stiffness results to be a simple spatial average of the local stiffness.

As the stiffness decreses due to increasing disorder also 
quantum phase fluctuations beyond
Gaussian level become relevant\cite{randeria01,benfatto01}. 
In general, in the presence of a constant gauge field $A$ 
the expansion of the energy per unit surface, up to fourth order can
be written as: 
\be
\lb{enexp}
\frac{E}{L^2}=\frac{ D_s}{2} A^2-\frac{Q a^2}{6} A^4
=\frac{J}{2}(\delta \theta)^2-\frac{J'}{24}
\xi_0^2(\delta \theta)^4
\ee
where $\delta \theta$ is the phase gradient in a given ($x$ or $y$) direction and 
we have restored explicitly the lattice spacing $a$. 
From Eq.\ \pref{enexp} it is evident that while $D_s$ provides the
stiffness the coupling $Q$ measures the anharmonicity of the phase fluctuations.
The second part of Eq. \pref{enexp} translates the expansion 
into an analogus expression for the phase 
gradient by means of the mininal-coupling substitution $A 
\rightarrow \delta \theta/2$.

In Eq.\ \pref{enexp} we also introduced the couplings $J\equiv D_s/4$ 
and $J'\equiv Q a^2/(4\xi_0^2)$ in order to make a closer analogy with the
$XY$ model, $H=J\sum_{\langle ij\rangle} (1-\cos(\theta_i-\theta_j))$
that is  usually considered as the prototype model for
phase fluctuations in a superconductor. The latter is defined in a
coarse-grained scale $\xi_0$ given by the coherence length. Notice
that contrary to $J$, the characteristic 
energy for anharmonic fluctuations $J'$ in Eq.~\pref{enexp} depends on
the coarse-grained scale $\xi_0$.   
For the $XY$ model the gradient expansion  is
given by Eq.~\pref{enexp} with 
$J'=J$.

Quantum corrections to $D_s$ can be computed with a  
perturbative approximation from Eq.\ \pref{enexp}:
\be
\lb{quantum}
  D_s=D_s^0\left[ 1-\frac{J'}{J}\frac{\xi_0^2}{2}\langle (\delta
  \theta)^2\rangle\right ]
\ee 
where $D_s^0$ is the (bare) value obtained by the Gaussian expansion
in Eq.\ \pref{enexp} and $\langle (\d \theta)^2\rangle$ is computed
including the dynamics of the phase (cf. e.g. \cite{benfatto01,noi}). 

%
%
%
%

The coherence
length has been determined from the 
decay of the gap amplitude correlations which yields $\xi_0\approx 1a$, 
so that for the present parameters $Q/D_s  \approx J'/J$.
The upper left inset to Fig.\ref{fig-ds} shows the ratio $Q/D_s$ as
obtained from the fit of $E(A)$ to Eq.~\pref{enexp}. 
For increasing disorder $Q/D_s  \approx J'/J$ becomes strongly
enhanced leading rapidly to a larger phase-fluctuation correction
within the Hubbard as compared to the $XY$ model where $J'/J=1$.
Above a critical value of the disorder strength $V_0/t \approx 3$ a  SC-I
transition is obtained when $D_s=0$.
Within an RPA analysis of the full model on small clusters
we have also checked 
\cite{seibold} that $\langle (\d \theta)^2\rangle$ is increasing
with disorder, however, the dependence on $V_0$ being much weaker
than that for $Q/D_s$.

The above discussion shows that the $XY$ model fails in general to describe
quantitatively corrections of the superfluid stiffness due
to numerical differences in the quartic coefficient. 
Notwithstanding a disordered version of it can help to get
some insight on the anomalously large increase of the $J'/J$ ratio as
a function of disorder. Let us consider for simplicity a
one-dimensional XY-model
in the presence of a constant gauge
field, $H=\sum_{i} J_{i,i+1}(1-\cos(\theta_i-\theta_{i+1}-2A))$. 
In order to describe a disordered model we consider a random distribution of
couplings $J_{i,i+1}$.
To derive the coefficients $J,J'$ of the expansion \pref{enexp} 
we can use the fact that in 1D the current in each link, $j(A)$, must be
conserved, so that $-J_{i,i+1} \sin (\theta_i-\theta_{i+1}-2A)=j(A)/2$.
%
%
By inverting this relation, summing over the index $i$ and expanding 
$j(A)/2\approx J A-\frac{J'}{6} A^3$ we then obtain that the
coefficients are given by the spatial averages, 
 $J^{-1}= \overline{ 1/J_{i,i+1}}$ (see also Ref.~\cite{par98}), and 
$J'=J^4\; \overline{ 1/J_{i,i+1}^3}$. 
By computing for example the ratio $J'/J$ for a Gaussian
distribution of local coupling values one can see that it 
increases very rapidly as the distribution width
increases, leading to a  non-vanishing occurrence probability for very low values of
$J_{i,i+1}$, which have a much 
stronger effect on $J'$ than on $J$.
As a byproduct this computation also shows the importance of phase
relaxation,  
without which $J$ would be defined as 
$J= \overline{J_{i,i+1}}$. 
Obviously, in 2D the current in each link is not constant. Thus in principle
the current would  try to avoid links with very low $J_{ij}$ couplings to
avoid coexistence of large phase change and large current, which from
 Fig. \ref{fig-curr} seems to be indeed the case.
However, these patterns have an almost one dimensional character 
so that the above argument could be still qualitatively correct even in 2D.
Nevertheless the variation of local $J_{ij}$ values along the 
active paths (i.e. paths with current)
can be smaller than along paths with vanishing current.

In conclusion, we have analyzed from a real space BdG approach
for the attractive Hubbard model, the reaction
of a strongly disordered SC system to an applied transverse vector potential.
The first main result concerns the standard decoupling  between transverse and longitudinal response which does not apply in the presence of disorder. 
We have evaluated the 
corrections of the superfluid stiffness 
to the dirty BCS limit  (cf. Fig. \ref{fig-ds}) and obtained a strong
reduction of the superfluid stiffness due to the relaxation of 
the order parameter phase for 
large disorder. We also find that the anharmonicity  of the phase 
fluctuations is increased by disorder. Beyond mean-field, 
this enhances the effect of quantum fluctuations in producing a 
zero-temperature transition to a non-SC phase. Both
effects can explain the deviations of the zero-temperature superfluid stiffness
from the BCS dirty limit reported recently for strongly disordered NbN
films\cite{mondal10}. 


Secondly, our calculations have revealed that the stiffness for strong
disorder is
dominated by quasi one-dimensional percolative paths 
along sites with large gap parameters.
In some sense this mimics the finding of Ref. \cite{IMFIM} where the phase
diagram of a model, similar to the present one in the large $U/t$ limit,
has been analyzed within the so-called cavity mean-field approximation.
Interestingly it was found that there is a regime of broken-replica symmetry
where the partition function is determined by a small number of 
paths. Taking that a similar glassy physics is at
work in our model, where the response is governed by a few number
of percolative paths, then this should also appear in the distribution
functions of order parameters and energy level spacings. 
An analysis of this issue is in progress.

\end{document}